\documentclass[RNASS]{aastex62}
\usepackage{amsmath}

\def\kpc{\,\mathrm{kpc}}
\def\muas{\,\mu\mathrm{as}}
\def\magnitude{\,\mathrm{mag}}

\graphicspath{{./}{figures/}}

\shorttitle{Distance estimates for Gaia RV stars}
\shortauthors{McMillan, P.J.}

\begin{document}

\title{Simple distance estimates for \textit{Gaia} DR2 stars with radial velocities}

\correspondingauthor{Paul McMillan}
\email{paul@astro.lu.se}

\author[0000-0002-8861-2620]{Paul J. McMillan}
\affil{Department of Astronomy and Theoretical Physics, Lund Observatory, Box 43, SE-221 00 Lund, Sweden}


\section*{} \label{sec:intro}

As noted by numerous authors since \cite{Stromberg27}, the inverse of a measured parallax is a biassed estimate of the distance to a star, increasingly so as the relative uncertainty becomes larger. There are essentially three competing effects that one must consider: 1) the volume at distances between $s$ and $s+\mathrm{d}s$ increases like $s^2$; 2) the true spatial distribution of stars is not uniform; and 3) selection effects:  the probability of a star at distance $s$ entering a catalogue varies with $s$ if there is any magnitude limit to the survey (because e.g., intrinsically faint stars become too faint to enter the catalogue). For a discussion see \citet{Luriea18}.

{\it Gaia} DR2 contains radial velocities (and therefore full 3D velocities) for 7 million of its 1.7 billion sources \citep{GaiaDR2:Summary,GaiaDR2:RV}. These stars are therefore of particular interest to people studying Milky Way dynamics. They are selected to have $3550 < T_{\rm eff} < 6900 \,\mathrm{K}$ and $G_{RVS} < 12\magnitude$, and the selection effect that applies for this subset of stars is quite different to that which applies to the complete {\it Gaia} sample. 

We estimate the distance ($s$) to these stars using the {\it Gaia} parallax $\varpi$ and $G_{RVS}$ magnitude \citep[calculated from $G$ and $G_{\rm RP}$ using the approximation from][]{GaiaDR2:Summary}, and making the Bayesian estimate
\[
P(s|\, \varpi,G_{RVS}) \propto \; P(\varpi | s,\sigma_\varpi) \times s^2 P(\mathbf{r}) P(M_{G_{RVS}}).
\]
where $M_{G_{RVS}}$ is the absolute magnitude in the $G_{RVS}$ band. We describe this distance pdf in terms of its expectation value and standard deviation. We assume that the uncertainty in $G_{RVS}$ is sufficiently small on the scale of the prior $P(M_{G_{RVS}})$ that it can be treated as a $\delta$-function. 
{\it Gaia} DR2 parallaxes have a zero-point offset, and systematic uncertainties. We attempt to compensate for these by `correcting' the parallax estimates for a zero-point of $-29\muas$ and adding an uncertainty of $42\muas$ in quadrature with the quoted value. This reflects values found from the analysis of quasars by \citet[table 4]{GaiaDR2:Astrometry}.

The density model used to give the prior on position $P(\mathbf{r})$ is the same as that used by \citet[henceforth PM18]{McMillanea18}, plus a bulge component taken from \citet{PJM:MassModels2}, normalised to give the same bulge-to-disc mass ratio.

\begin{figure}
\plotone{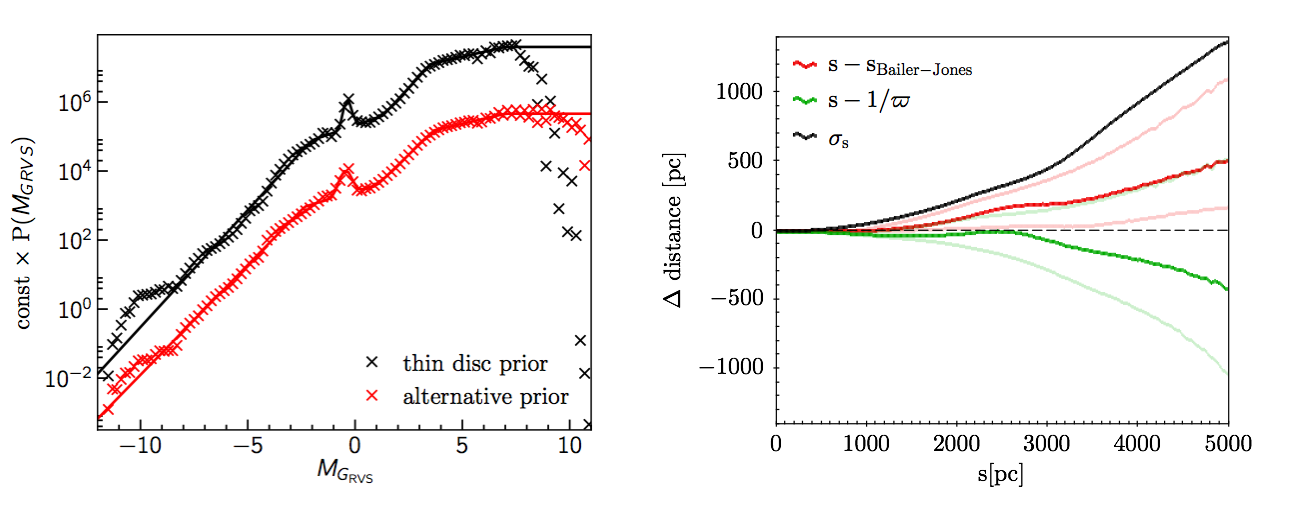}
\caption{Left: Prior used (black) and alternative prior - we show the data from the model population, and the functions fit to them. Right: Comparison to \protect\citet{BailerJonesea18} distances or na\"{i}ve $1/\varpi$ estimate (in both cases, dark line is the median, faint lines the 16th \& 84th percentiles) as a function of our distance estimate. Our median uncertainty $\sigma_s$ is also shown.\label{fig:PMGRVS}}
\end{figure}

Our prior $P(M_{G_{RVS}})$ is an approximation to the distribution of $M_{G_{RVS}}$ one would expect for stars with $3550< T_{\rm eff}< 6900\,\mathrm{K}$, under sensible assumptions about the underlying population, following PM18. We assume that the population has the same IMF and age distribution as in the `Age' prior used by PM18, and because most of these stars will belong to the thin disc, we take a thin-disc-like prior in metallicity ($P([M/H])$ is a Gaussian with mean $-0.1$ and dispersion $0.3$). We model this with PARSEC isochrones \citep{Marigoea17}; the results are shown in Figure~\ref{fig:PMGRVS}. Our prior is a piecewise linear fit to this in $\log (P(M_{G_{RVS}}))$  plus a Gaussian red clump component, also shown. 
Since we are not using colour information to derive the distances, this broad approximation is reasonable, though clearly more precise methods can be used for these data. We have explored the effect of using an alternative prior -- derived from a simulated population with flat metallicity and age distributions (also shown in Figure~\ref{fig:PMGRVS}). The effects are small, on average a distance increase of $\sim2$\% with the flat prior to a distance of about $s=4\kpc$, with the 1$\sigma$ differences being $\sim[-1,+3]$\% at $4\kpc$.


\citet{BailerJonesea18} used {\it Gaia} parallaxes, and a prior based on the expected distribution of \emph{all} stars in the {\it Gaia} catalogue to derive distance estimates for 1.3 billion stars. Our study differs because it is specifically designed to apply \emph{only} to the stars with published {\it Gaia} DR2 radial velocities, which means we use the $G_{\rm RVS}$ magnitude as input (ensuring we consider the relevant selection effects for this sample). Figure~\ref{fig:PMGRVS} shows a comparison between our distances and those from \citeauthor{BailerJonesea18}. Our estimates diverge at larger distances as the uncertainties become more significant. We note that, by accident of the selection function, the na\"{i}ve estimate $1/\varpi$ performs well as a distance estimate out to several $\kpc$ -- see \citet{SchoenrichAumer17} for a similar result.

We note the important caveat that we have not considered dust extinction, though this is much less important than it would be if we had used colour information. 
The effects will be important in more extinguished regions.

Our code is available through \url{https://github.com/PaulMcMillan-Astro/GaiaRVStarDistances} \citep{zenodo.1270548}. The distance estimates can be found at \url{http://www.astro.lu.se/~paul/GaiaDR2_RV_star_distance.csv.gz} and archived at \url{https://doi.org/10.5281/zenodo.1268353}. They have already been used by \cite{Quillenea18} to study 
velocity substructure.


\section*{Acknowledgements}
I am grateful to I. Carrillo and A. Quillen who encouraged publication. I am supported by the Swedish National Space Board, and the Swedish research council.
This work has made use of data from the European Space Agency
mission {\it Gaia} (\url{https://www.cosmos.esa.int/gaia}), processed by
the {\it Gaia} Data Processing and Analysis Consortium (DPAC,
\url{https://www.cosmos.esa.int/web/gaia/dpac/consortium}). 

\software{
	Matplotlib \citep{Matplotlib}
}



\begin{thebibliography}{}


\bibitem[Bailer-Jones et al.(2018)]{BailerJonesea18} Bailer-Jones, C.,  et al.\ 2018 , arXiv:1804.10121.

\bibitem[Gaia Collaboration et al.(2018)]{GaiaDR2:Summary}
{Gaia Collaboration} et al. arXiv:1804.09365

\bibitem[Hunter (2007)]{Matplotlib} Hunter, J.\ 2007, Computing In Science \& Engineering, 9, 3


\bibitem[Katz et al.(2018)]{GaiaDR2:RV} Katz, D. et al.\ 2018, arXiv:1804.09372.

\bibitem[Lindegren et al.(2018)]{GaiaDR2:Astrometry} {Lindegren L.}, {et al.}, preprint, arXiv:1804.09366

\bibitem[Luri et al.(2018)]{Luriea18} Luri, X. et al.\ 2018, arXiv:1804.09376.

\bibitem[Marigo et al.(2017)]{Marigoea17} Marigo, P. et al.\ 2017, \apj, 835, 77.
\bibitem[McMillan et al.(2018)]{McMillanea18} McMillan, P.~J. et al.\ 2018, \mnras, 1016. [PM18]
\bibitem[McMillan(2017)]{PJM:MassModels2} McMillan, P.\ 2017, \mnras, 465, 76.

\bibitem[McMillan(2018)]{zenodo.1270548} McMillan, P.\ 2018, PaulMcMillan-Astro/GaiaRVStarDistances: Publication release, v0.1, Zenodo, doi:10.5281/zenodo.1270548


\bibitem[Quillen et al.(2018)]{Quillenea18} Quillen, A. et al.\ 2018,  arXiv:1805.10236.



\bibitem[\protect\citeauthoryear{Sch{\"o}nrich \& Aumer}{2017}]{SchoenrichAumer17} Sch{\"o}nrich R., Aumer M., 2017, MNRAS, 472, 3979

\bibitem[\protect\citeauthoryear{{Str{\"o}mberg}}{{Str{\"o}mberg}}{1927}]{Stromberg27}
{Str{\"o}mberg} G., \apj,   1927, {65, 238}




\end{thebibliography}
\end{document}